\documentclass[galaxies,article,accept,oneauthor,pdftex,12pt,a4paper]{mdpi}

\lastpage{x}
\doinum{10.3390/------}
\pubvolume{xx}
\pubyear{2014}
\history{Received: xx / Accepted: xx / Published: xx}
\pdfoutput=1

\usepackage{amsfonts} 
\usepackage{amstext}
\usepackage{amssymb} 
\usepackage{subfig}

\newcommand{\cvg}{$d_{R}V_{0}\;$}
\newcommand{\cvgA}{$d_{R}V_{0}$}

\Title{The inner regions of disk galaxies: a constant baryonic fraction?}

\Author{Federico Lelli $^{1}$}

\address{$^{1}$ Department of Astronomy, Case Western Reserve University,
10090 Euclid Ave, Cleveland, OH 44106, USA; E-mail: federico.lelli@case.edu;
Tel.: +1-216-368-0917}

\abstract{For disk galaxies (spirals and irregulars), the inner
circular-velocity gradient \cvg (inner steepness of the rotation
curve) correlates with the central surface brightness $\Sigma_{*, 0}$
with a slope of $\sim$0.5. This implies that the central dynamical
mass density scales almost linearly with the central baryonic
density. Here I show that this empirical relation is consistent
with a simple model where the central baryonic fraction $f_{\rm bar,
\, 0}$ is fixed to 1 (no dark matter) and the observed scatter is
due to differences in the baryonic mass-to-light ratio $M_{\rm
bar}/L_{\rm R}$ (ranging from  1 to 3 in the $R$-band) and in the
characteristic thickness of the central stellar component $\Delta z$
(ranging from 100 to 500~pc). Models with lower baryonic fractions
are possible, although they require some fine-tuning in the values
of $M_{\rm bar}/L_{\rm R}$ and $\Delta z$. Regardless of the actual
value of $f_{\rm bar, 0}$, the fact that different types of galaxies
do not show strong variations in $f_{\rm bar, 0}$ is surprising,
and may represent a challenge for models of galaxy formation in
a $\Lambda$CDM cosmology.}


\begin{document}
 
\section{Introduction}\label{sec:intro}

Galaxies are known to follow tight scaling relations linking their observed
baryonic content to their dynamical properties. Pressure-supported systems
(ellipticals, bulges, and dwarf spheroidals) follow the Faber-Jackson
relation \cite{Faber1976, Sanders2010}, which links the total luminosity
of the system (proxy for the baryonic mass) to its mean velocity dispersion
(proxy for the dynamical mass). Rotation-supported systems (lenticulars,
spirals, and dwarf irregulars) follow the baryonic Tully-Fisher relation
(BTFR) \cite{McGaugh2000}, which links the total baryonic mass $M_{\rm bar}$
(gas plus stars) to the asymptotic velocity along the flat part of the
rotation curve $V_{\rm flat}$ \cite[e.g.][]{Verheijen2001b, McGaugh2005,
Noordermeer2007b}. While $V_{\rm flat}$ is related to the total dynamical
mass of the galaxy, the inner steepness of the rotation curve provides
information on the central dynamical mass density, including both baryons
and dark matter (DM). Early studies \cite{Kent1987, Corradi1990, Casertano1991,
Broeils1992} pointed out that, in disk galaxies, the shape of the luminosity
profile and the shape of the rotation curve are related, suggesting a close
link between the distribution of baryons and the distribution of the dynamical
mass (luminous and/or dark). This has been confirmed by several subsequent
studies \cite{deBlok1996, Tully1997, Marquez1999, Noordermeer2007c, Swaters2009,
Lelli2010}, which have substantially increased the number of galaxies with
high-quality rotation curves, spanning the Hubble sequence from lenticulars
to irregulars (Irrs). The observational evidence is concisely summarized by
the so-called ``Renzo's rule'' \cite{Sancisi2004}: for any feature in the
luminosity profile of a galaxy there is a corresponding feature in the
rotation curve, and vice versa.

\begin{figure}[h!]
\centering
\includegraphics[width=\textwidth]{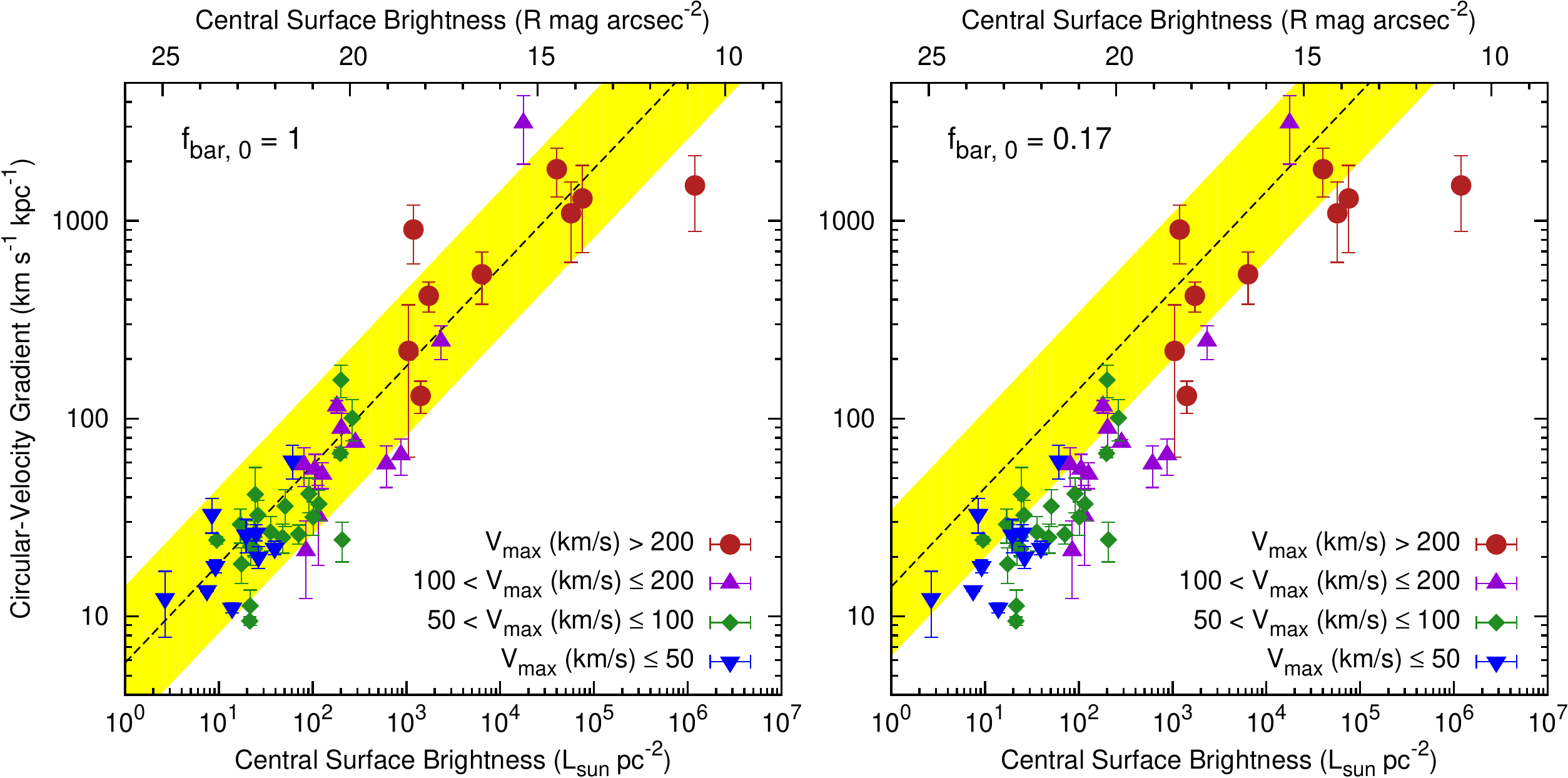}
\caption{The $d_{R}V_{0} - \Sigma_{*, 0}$ scaling relation. Galaxies are
coded by the value of the maximum velocity $V_{\rm max}$ observed along the
rotation curve. Galaxies with $\Sigma_{*, 0}\gtrsim 10^{3}$~L$_{\odot}$~pc$^{-2}$
are typically dominated by a bulge component in the inner parts and have
more uncertain values of $d_{R}V_{0}$ \cite[see][for details]{Lelli2013}.
The yellow band shows a model where the central baryonic fraction is constant
for each galaxy ($f_{\rm bar, 0}=1$ in the left panel, $f_{\rm bar, 0} = 0.17$
in the right one); the width of the band considers that the value of $M_{\rm bar}
/L_{\rm R}$ can vary from 1 to 3, $\Delta z$ from 100 to 500 pc, and $\alpha$
from 0.5 to 1. The dashed line indicates an ``intermediate'' model with
$M_{\rm bar}/L_{\rm R}=2$, $\Delta z = 300$~pc, and $\alpha = 0.75$.}
\label{fig:Gradient}
\end{figure}

In \cite{Lelli2013} we measured the inner circular-velocity gradient \cvg
for a sample of 52 galaxies with high-quality rotation curves, ranging
from bulge-dominated galaxies (S0 to Sb) to disk-dominated ones (Sc to Irrs).
\cvg is defined as $dV/dR$ for $R \rightarrow 0$, thus measuring the inner
steepness of the galaxy rotation curve. We found that \cvg correlates with
the central surface brightness $\Sigma_{*, 0}$ over more than two orders
of magnitude in \cvg and four orders of magnitude in $\Sigma_{*, 0}$ (see
Fig. \ref{fig:Gradient}). This is a scaling relation for rotation-supported
systems, which is analogous to the BTFR for the innermost regions of disk
galaxies. These two empirical laws imply that, to a first approximation,
the baryonic properties of a galaxy can predict the shape of the rotation
curve, and vice versa. For example, the rotation curves of late-type
galaxies can be roughly described by a nearly solid-body rise in the
inner regions (given by \cvgA) and the flattening in the outer parts
($V_{\rm flat}$) \cite[e.g.][]{Begeman1987, Swaters2009}. Thus, if the
central surface brightness and the total baryonic mass of a galaxy are
known, one can approximately predict the shape of its rotation curve using
the $d_{R}V_{0}-\Sigma_{*, 0}$ and the $V_{\rm flat}-M_{\rm bar}$ relations.
The predictive power of baryons is an empirical fact, independent of any
underlying theory. This is quite surprising in a DM-dominated Universe.

In this paper I build a toy model that naturally reproduces the slope,
normalization, and scatter of the $d_{R}V_{0} - \Sigma_{*, 0}$ relation.
I also discuss the general implications that this relation poses to the
baryonic fraction in the innermost regions of disk galaxies.

\section{A toy model for the $d_{R}V_{0} - \Sigma_{*, 0}$ relation}\label{sec:baryonFrac}

For a general 3D distribution of mass, the circular velocity $V$
of a test particle orbiting at radius $R$ is given, to a first
approximation, by
\begin{equation}\label{eq:Newton}
\dfrac{V^{2}}{R} = \alpha \, \dfrac{G M_{\rm{dyn}}} {R^{2}} =
\dfrac{4}{3} \pi \, \alpha \, G \, \overline{\rho}_{\rm{dyn}} \, R,
\end{equation}
where $G$ is Newton's constant, $M_{\rm{dyn}}$ is the dynamical mass
within $R$, $\overline{\rho}_{\rm{dyn}} = M_{\rm{dyn}}/ \frac{4}{3}
\pi R^{3}$ is the mean dynamical mass density within $R$, and $\alpha$
is a factor of the order of unity that depends on the 3D distribution
of mass. For a spherical mass distribution, the Newton's theorem gives
$\alpha$=1 \cite[e.g.][]{BT94}. For a thin exponential disk with scale
length $R_{\rm d}$, $\alpha$ varies with radius \cite[cf.][]{Freeman1970}:
$\alpha \simeq 1$ at $R=R_{\rm d}$ and monotonically decreases for
$R \lesssim R_{\rm d}$ ($\alpha\simeq 0.75$ at $R = 0.5 \, R_{\rm d}$;
$\alpha\simeq 0.5$ at $R = 0.25 \, R_{\rm d}$). The surface brightness
profiles of disk galaxies often deviate from a pure exponential in the
inner regions: if the luminosity profile shows an inner ``flattening'',
as it is often observed in dwarf galaxies \cite[e.g.][]{Swaters2002b},
the value of $\alpha$ near the center decreases with respect to an
exponential disk. For these reasons, I consider that $\alpha$ can vary
between 0.5 and 1 in the inner parts of different galaxies ($R\lesssim
0.5 R_{\rm d}$).

If we assume that the dynamical mass density converges to a finite
value $\rho_{\rm{dyn, 0}}$ towards the center (as is reasonable
for actual galaxies), in the limit $R \to 0$ we have
\begin{equation}\label{eq:baryonFrac}
\dfrac{dV}{dR} \simeq \dfrac{V}{R} = \sqrt{\dfrac{4}{3} \pi \, \alpha G \rho_{\rm{dyn, 0}}} =
\sqrt{\dfrac{4}{3} \pi \, \alpha G \dfrac{\rho_{\rm{bar, 0}}}{f_{\rm{bar, 0}}}},
\end{equation}
where $\rho_{\rm{bar, 0}}$ is the central baryonic mass density and
$f_{\rm{bar, 0}} = \rho_{\rm{bar, 0}}/\rho_{\rm{dyn, 0}}$ is the
central baryonic fraction. Note that $f_{\rm{bar, 0}}$ may strongly
differ from the ``cosmic'' baryonic fraction (= 0.17) given by the
Cosmic Microwave Background \cite{Komatsu2009} and observed in galaxy
clusters \cite{McGaugh2010}. In a $\Lambda$CDM cosmology, $f_{\rm{bar,
0}}$ is determined by the complex formation and evolution history of
the object, involving galaxy mergers, gas inflows, star formation,
stellar and AGN feedback, etc. Thus, we expect $f_{\rm{bar, 0}}$ to
vary from galaxy to galaxy, possibly from 0.17 (the cosmic value)
up to 1 (baryon dominance). Moreover, in a given galaxy, the baryonic
fraction $f_{\rm bar}$ can vary with radius up to a factor of 10
\cite[see e.g.][]{McGaugh2004} due to the relative contributions
of baryons and DM to each point of the rotation curve. The baryonic
fractions deduced from Eq.~\ref{eq:baryonFrac} are formal extrapolations
for $R \to 0$, but in practice they are representative of the innermost
galaxy regions accessible by the available rotation curves (typically
for $R \lesssim 0.5 R_{\rm d}$, see \cite{Lelli2013} for details).

Observationally, we measure either $\mu_{0}$ (in units of mag
arcsec$^{-2}$) or $\Sigma_{*, 0}$ (in units of $L_{\odot}$~pc$^{-2}$).
The latter is given by
\begin{equation}
 \Sigma_{*, 0} = \int_{-\infty}^{\infty} \dfrac{\rho_{\rm bar, 0}}{M_{\rm{bar}}/L} dz
 \simeq \dfrac{\rho_{\rm bar, 0}}{M_{\rm{bar}}/L} 2 \Delta z,
\end{equation}
where $\Delta z$ is the characteristic thickness of the stellar
component in the central regions (either a disk, a bulge, a
bar/pseudo-bulge, or a nuclear star cluster) and $M_{\rm{bar}}/L$
is the baryonic mass-to-light ratio, including molecules and
other undetected baryons (the atomic gas density is generally
$<$10 M$_{\odot}$~pc$^{-2}$ in the inner regions and, thus,
negligible to a first approximation). Therefore, we expect
the following relation
\begin{equation}\label{eq:Newt}
 d_R V_{0} = \sqrt{\dfrac{2}{3} \pi \, \alpha G \dfrac{M_{\rm bar}/L}{\Delta z \, f_{\rm bar, 0}}} \, \sqrt{\Sigma_{*,0}}.
\end{equation}

Remarkably, a least-square fit to the data-points in Fig.~\ref{fig:Gradient}
returns a slope of $\sim$0.5 (equivalent to $-0.2$ when the central surface
brightness is expressed in units of mag~arcsec$^{-2}$ instead of $L_{\odot}$
pc$^{-2}$, see \cite{Lelli2013, Lelli2014a}). The actual value of the slope
remains uncertain due to several effects in the determination of \cvg and
$\Sigma_{*, 0}$, but it can be constrained between $\sim$0.4 and $\sim$0.6.
Despite these uncertainties, it is interesting to check whether the zero
point and the observed scatter along the relation are consistent with typical
values of $\alpha$, $\Delta z$, $M_{\rm bar}/L_{\rm R}$, and $f_{\rm bar, 0}$.
In both panels of Fig.~\ref{fig:Gradient}, the yellow band is calculated
assuming that $f_{\rm bar,0}$ is \textit{constant} for each galaxy ($f_{\rm
bar, 0} = 1$ in the left panel, $f_{\rm bar, 0} = 0.17$ in the right one),
while $\alpha$ varies from 0.5 to 1, $\Delta z$ varies from 100 to 500 pc,
and $M_{\rm bar}/L_{\rm R}$ varies from 1 to 3. The dashed line shows an
``intermediate'' model with $\alpha=0.75$, $\Delta z = 300$~pc, and $M_{\rm
bar}/L_{\rm R} =2$. The width of the band is dominated by the variation
in $\Delta z$ (factor $\sqrt{5}$) followed by the variations in $M_{\rm bar}
/L_{\rm R}$ (factor $\sqrt{3}$) and $\alpha$ (factor $\sqrt{2}$). Note that
the objects in the sample cover a wide range in total mass ($20 \lesssim
V_{\rm max}\lesssim 300$ km~s$^{-1}$, as indicated by the different
symbols in Fig.~\ref{fig:Gradient}), and span the entire Hubble sequence
going from bulge-dominated galaxies (S0 to Sb), typically characterized by
old stellar populations in the central regions, to disk-dominated galaxies
(Sc to Irr), characterized by young stellar populations. Considering possible
differences in their star formation history, molecular content, metallicity,
internal extinction, and initial mass function, a variation in $M_{\rm bar}
/L_{\rm R}$ by a factor of 3 is a rather conservative choice \cite[cf.]
[]{Bell2001, Portinari2004, Into2013}.

In the left panel of Fig.~\ref{fig:Gradient}, I consider $f_{\rm bar, 0}= 1$
(no DM). The agreement with the observations is striking. The scatter along
the relation can be fully explained by variations in the 3D distribution
of baryons ($\Delta z$ and $\alpha$) and stellar populations ($M_{\rm bar}
/L_{\rm R}$), without any need of DM in the innermost galaxy regions
(typically within $\sim$0.5~$R_{\rm d}$). In the right panel, I consider
the opposite, extreme case where $f_{\rm bar, 0}$ is fixed to 0.17 (the
cosmic value). Clearly, this low baryonic fraction cannot reproduce the
observed relation, unless one significantly increases $\Delta z$ and/or
decreases $M_{\rm bar}/L_{\rm R}$. The characteristic thickness of the
central stellar component should be increased up to values of $\sim$3~kpc,
that appear quite unrealistic. The baryonic mass-to-light ratio, instead,
should be decreased down to $\sim0.1-0.2$ which is inconsistent with
standard stellar population synthesis models \cite[e.g.][]{Bell2001,
Portinari2004, Into2013}. Models with intermediate values of $f_{\rm bar, 0}$
are possible, provided that $M_{\rm bar}/L_{\rm R}$ and/or $\Delta z$ are
properly fine-tuned. For example, a plausible model is obtained by fixing
$f_{\rm bar, 0}\simeq 0.5$ ($\sim$3 times the cosmic value) and varying
$M_{\rm bar}/L_{\rm R}$ between 0.5 and 1.5, $\Delta z$ between 100 and
500~pc, and $\alpha$ between 0.5 and 1. Such a model cannot be distinguished
from the one in Fig.~\ref{fig:Gradient} (left panel) due to the degeneracies
between $M_{\rm bar}/L_{\rm R}$, $\Delta z$, and $f_{\rm bar, 0}$. Surface
photometry in IR bands ($K$-band or 3.6 $\mu$m) may improve the situation
and help to break these degeneracies, given that one expects a smaller
scatter in the values of $M_{\rm bar}/L$ \cite[e.g.][]{Verheijen2001b,
Bell2001}, which may translate into a smaller scatter around the $d_{R}V_{0}
-\Sigma_{*,0}$ relation. It would also be useful to investigate whether
the residuals around the central relation correlate with some galaxy
properties. For example, if baryons dominate the central galaxy regions
($f_{\rm bar,0}\simeq1$), one may expect the residuals to correlate with
the central colors (tracing $M_{\rm bar}/L_{\rm R}$) and/or the central
morphology (roughly tracing $\alpha$ and $\Delta z$). These issues will
be addressed in future studies.

\section{Discussion \& Conclusions}

I investigated the implications that the $d_{R}V_{0}-\Sigma_{*,0}$ relation
poses to the central baryonic fraction $f_{\rm bar, 0}$ of disk galaxies
(spirals and irregulars). Surprisingly, I found that the observed relation
is consistent with a model where $f_{\rm bar, 0}$ is fixed to 1 (no DM) and
the scatter is entirely given by variations in the baryonic mass-to-light ratio
$M_{\rm bar}/L_{\rm R}$ (from 1 to 3 in the $R$-band), in the characteristic
thickness of the central stellar component $\Delta z$ (from 100 to 500 pc),
and in the 3D shape of the gravitational potential (parametrized by $\alpha
\simeq 0.5$ to 1). Models with very low values of $f_{\rm bar, 0}$ (such as
the ``cosmic'' value of 0.17) are very unlikely, since they would require
values of $M_{\rm bar}/L_{\rm R}$ and $\Delta z$ that are in tension with
our current knowledge on stellar populations and galaxy structure. Models
with intermediate values of $f_{\rm bar, 0}$ ($\simeq$0.5) are possible,
provided that the parameters $M_{\rm bar}/L_{\rm R}$ and $\Delta z$ are
properly fine-tuned.

If $f_{\rm bar, 0}\simeq1$ and, thus, baryons dominate the innermost regions
of disk galaxies ($R \lesssim 0.5 R_{\rm d}$), even in low-luminosity and
low-surface-brightness ones, the whole controversy about cuspy \textit{versus}
cored DM density profiles is undermined \cite[see e.g.][]{Sancisi2004}. Cored
DM profiles may be allowed if their characteristic central surface density is
relatively low ($<$100 M$_{\odot}$~pc$^{-2}$), whereas cuspy DM profiles
would generally be disfavoured as they should have low concentrations that are
unexpected in a $\Lambda$CDM cosmology \cite[e.g.][]{McGaugh2007}. Broadly
speaking, a dominant baryonic component in the innermost galaxy regions would
leave ``little room'' for a central DM cusp. If $f_{\rm bar}<1$, both cored
and cuspy DM profiles may be allowed, but one should explain why, for galaxies
of very different masses and rotation velocities, the central baryonic density
scales almost linearly with the central dynamical mass density due to the
dominant DM halo.

Regardless of the actual value of $f_{\rm bar, 0}$, it is surprising that
models with fixed $f_{\rm bar, 0}$ can naturally reproduce the $d_{R}V_{0}
-\Sigma_{*,0}$ relation. In a $\Lambda$CDM cosmology, the central baryonic
fractions of galaxies are the result of complex baryonic physics (including
mergers, gas inflows, star formation, stellar and AGN feedback, etc.), which
likely depends on the specific properties of the galaxy (total mass, angular
momentum, environment, etc.). It is puzzling, therefore, that the data can
be accurately described using a fixed value of $f_{\rm bar, 0}$ for very
different types of galaxies, ranging from dwarf irregulars with $V_{\rm max}
\simeq 20-30$ km~s$^{-1}$ to bulge-dominated spirals with $V_{\rm max}
\simeq 200-300$ km~s$^{-1}$. A large variation in $f_{\rm bar, 0}$ from
galaxy to galaxy would introduce further scatter along the relation that
is not observed. Intriguingly, the situation is very different for the
BTFR. The zero point and slope of the BTFR imply that the global baryonic
fraction in the disk $f_{\rm d}$ is \textit{smaller} than the cosmic value
by $\sim$1-2 orders of magnitude and \textit{systematically decreases} with
$V_{\rm flat}$ \cite{McGaugh2010, McGaugh2012}. In other words, galaxies
must progressively lose more and more baryons during their formation with
decreasing mass (as suggested by the BTFR), but the fraction of baryons
in the inner regions should remain \textit{higher} than the cosmic value
and \textit{almost constant} in any type of galaxy (as suggested by the
$d_{R}V_{0}-\Sigma_{*,0}$ relation). This may be a challenge for models
of galaxy formation in a $\Lambda$CDM cosmology. I stress, however, that
the current observations constrain the slope of the $d_{R}V_{0}-\Sigma_{*,0}$
relation between $\sim$0.4 and $\sim$0.6 \cite{Lelli2013}: values slightly
lower/higher than 0.5 may point to systematic variations of $f_{\rm bar, 0}$
with $\Sigma_{*, 0}$. Future observational studies, therefore, should aim
at obtaining a better calibration of this scaling law.

\acknowledgements{Acknowledgements}

I thank the two anonymous referees for their constructive reports. I am
grateful to Filippo Fraternali, Stacy McGaugh, and Marcel Pawlowski for
stimulating discussions, helpful comments, and suggestions.



\bibliography{bibliography.bib}
\bibliographystyle{mdpi}

\end{document}